\documentclass[letterpaper]{JHEP3}
\usepackage{amsmath}
\usepackage{dsfont}
\usepackage{amssymb}
\usepackage{bm}
\newcommand{\eq}{\begin{equation}}
\newcommand{\eeq}{\end{equation}}

\title{Thermodynamics of dyonic black holes with Thurston horizon geometries}
\author{Mois\'es Bravo Gaete\\
   Facultad de Ciencias B\'asicas, Universidad
Cat\'olica del Maule, Casilla 617, Talca, Chile.\\
    E-mail: \email{mbravo-at-ucm.cl}}

\author{Mokhtar Hassa\"ine\\
Instituto de Matem\'atica y Fisica, Universidad de Talca,
Casilla 747, Talca, Chile.\\
E-mail: \email{hassaine-at-inst-mat.utalca.cl}}

\abstract{In five dimensions, we consider a model described by the
Einstein gravity with a source given by a scalar field and various
Abelian gauge fields with dilatonic-like couplings. For this model,
we are able to construct two dyonic black holes whose
three-dimensional horizons are modeled by two nontrivial homogeneous
 Thurston's geometries. The dyonic solutions are of
Lifshitz type with an arbitrary value of the dynamical exponent. In
fact, the first gauge field ensures the anisotropy asymptotic while
the remaining Abelian fields sustain the electric and magnetic
charges. Using the Hamiltonian formalism, the mass, the electric and
magnetic charges are explicitly computed. Interestingly enough, the
dyonic solutions behave like Chern-Simons vortices in the sense that
their electric and magnetic charges turn to be proportional. The
extension with an hyperscaling violating factor is also scrutinized
where we notice that for specific values of the violating factor,
purely magnetic solutions are possible. }

\begin{document}
%%%%%%%%%%%%%%%%%%%%%%%
\section{Introduction}
%%%%%%%%%%%%%%%%%%%%%%%
During the last decade, some promising efforts have been made to
extend the standard adS/CFT correspondence to new areas of physics,
and more particulary to physical systems enjoying an anisotropy
symmetry. By anisotropy, we mean that the space and the time are
allowed to scale with different weights. In this optics, the pioneer
works were done in the context of physical models invariant under
the Galilean-Schr\"odinger symmetry \cite{Son:2008ye,
Balasubramanian:2008dm}, see also \cite{Duval:2008jg} for a
geometric approach. Soon after, it was realized that similar
holographic considerations can also be translated to the case of
scale invariant Lifshitz fixed point systems without Galilean
invariance. In this case, the gravity dual metric is commonly known
as the Lifshitz spacetime \cite{Kachru:2008yh} and its
representative metric in arbitrary $D$ dimension can be
parameterized as
\begin{eqnarray}
ds^2=-r^{2z}dt^2+\frac{dr^2}{r^2}+r^2\sum_{i=1}^{D-2}dx_i^2.
\label{Lifshitzm}
\end{eqnarray}
In order to avoid as well as possible cumbersome formulas, we have
chosen to take the adS radius $l=1$. It is simple to see that the
anisotropic transformations defined by
\begin{eqnarray}
t\to\lambda^z\,t,\qquad r\to\frac{1}{\lambda}\,r,\qquad
x_i\to\lambda\, x_i, \label{aniTransf}
\end{eqnarray}
are part of the isometry of the Lifshitz metric. Here the constant
$z$ which reflects the anisotropy is called the dynamical exponent.
In analogy with the adS case, black holes with  a Lifshitz
asymptotic (\ref{Lifshitzm}), the so-called {\it Lifshitz black
holes}, would also have a certain interest for holographic
considerations. This interest has grew up during the last time as
shown by the important literature on the subject, see e. g.
\cite{Taylor:2008tg, AyonBeato:2009nh, Pang:2009pd, Maeda:2011jj,
Zangeneh:2016cbv, Quinta:2016eql, Ayon-Beato:2015jga,
Bravo-Gaete:2015xea, Zangeneh:2015uwa, Correa:2014ika, Fan:2015yza,
Hendi:2013zba, Alvarez:2014pra,Bravo-Gaete:2013dca}. From these
different examples it is clear that, in contrast with the adS
isotropic case, the Einstein-Hilbert action with eventually a
cosmological constant is not enough to sustain the Lifshitz metric.
In fact, in standard gravity, Lifshitz black holes can only exist
provided  the introduction of some extra matter fields
\cite{Pang:2009pd,Maeda:2011jj, Zangeneh:2016cbv} while higher-order
gravity theories with or without matter source may also source the
Lifshitz spacetimes, see e. g. \cite{AyonBeato:2009nh,
Ayon-Beato:2015jga, Bravo-Gaete:2015xea}.

Before proceeding, we would like first to enlarge the notion of
Lifshitz black holes. In its standard form, the $(D-2)-$dimensional
base manifold of the Lifshitz metric (\ref{Lifshitzm}) is an
Euclidean flat space. This restriction on the manifold ensures that
the isometry group of the standard Lifshitz metric (\ref{Lifshitzm})
contains in addition to the anisotropic transformations
(\ref{aniTransf}), the spacetime translations $x_i\to x_i+c_i$ and
$t\to t+t_0$ as well as the spatial rotation $\vec{x}\to R\,\vec{x}$
with $R\in \mbox{SO}(D-2)$. The algebra of the corresponding
generators or equivalently of the Killing vector fields form the
so-called Lifshitz algebra. Nevertheless, there also exist black
hole solutions with a non-flat base manifold, see e. g.
\cite{Cadeau:2000tj,Mann:2009yx, Matulich:2011ct,
Olivares:2013uha,Hassaine:2015ifa}, whose asymptotic resembles the
Lifshitz one but with a different base manifold
\begin{eqnarray}
ds^2=-r^{2z}dt^2+\frac{dr^2}{r^2}+\sum_{i,j=1}^{D-2}g_{ij}(x,r)dx_i
dx_j. \label{Lifshitznonst}
\end{eqnarray}
%The choice of the topology of the $(D-2)-$dimensional event horizon
%may be as arbitrary as one escapes from the hypothesis of the
%Hawking's theorem. Nevertheless, in the present case
For such a metric of course, the isometry group will explicitly
depend on the form of the transverse metric $g_{ij}$. Note that the
isometry group of metric-like (\ref{Lifshitznonst}) may not contain
the anisotropic dilatations as it occurs for a spherical or
hyperboloid transverse metric, \cite{Mann:2009yx,Tarrio:2011de}.
There also examples of black hole solutions whose asymptotic forms
match with (\ref{Lifshitznonst}) with more than one anisotropic
direction \cite{Cadeau:2000tj,Hassaine:2015ifa}. In these cases, the
standard dilatation transformations (\ref{aniTransf}) are
generalized to
\begin{eqnarray}
t\to\lambda^z\,t,\qquad r\to\frac{1}{\lambda}\,r,\qquad
x_i\to\lambda^{\alpha_i} \,x_i, \label{aniTransgener}
\end{eqnarray}
where the coordinates $x_i$ for which $\alpha_i\not=1$ represent the
additional anisotropic directions. We find then appropriate to
extend the terminology of Lifshitz black holes to black hole
spacetime whose asymptotic metric mimics (\ref{Lifshitznonst}) and
which is invariant {\it at least} under the general dilatation
transformations (\ref{aniTransgener}). In other words, we only
demand that the isometry group  contains at least the dilatation
generator associated to (\ref{aniTransgener}) as well as the
generator of time translation. There is a certain interest in
extending the notion of Lifshitz black holes as we have done.
Indeed, Lifshitz black holes as defined by
(\ref{Lifshitznonst}-\ref{aniTransgener}) have been shown to exit in
the case of standard General Relativity for dimensions greater than
five \cite{Cadeau:2000tj,Hervik:2003vx}. The restriction on the
dimension, namely $D\geq 5$, results from the fact that the
horizon's topologies of these generalized Lifshitz solutions are
modeled by some of the Thurston's geometries \cite{Thurston} which
can only be defined for dimensions greater than three. In fact, as
conjectured by Thurston and later on proved by Perelman, any compact
orientable three-dimensional Riemannian manifold can be modeled by
one of the eight Thurston's geometries\footnote{In fact, these eight
three-dimensional Thurston's geometries can be extended in
dimensions $D>3$.} which are the Euclidean space $\mathbb{E}^{3}$,
the three-sphere $\mathbb{S}^{3}$, the hyperbolic space
$\mathbb{H}^{3}$, the products $\mathbb{S}^{1} \times
\mathbb{S}^{2}$ and $\mathbb{S}^{1} \times \mathbb{H}^{2}$. In
addition, there exist three other possible geometries which are
neither of constant curvature nor of product of constant manifolds,
called the {\it Nil} geometry, the {\it Solv} geometry and the
geometry of the universal cover of $SL_{2}(\mathbb{R})$. These
exotic geometries  have the following representative metrics
\begin{subequations}
\begin{eqnarray}
\label{Solvg}
&&\mbox{Solv}:\qquad d\tilde{s}^2=x_3^{2}dx_1^2+\frac{1}{x_3^{2}}dx_2^2+\frac{1}{x_3^{2}} dx_3^2,\\
\nonumber\\
\label{Nilg}
&&\mbox{Nil}:\qquad d\tilde{s}^2=dx_1^2+dx_2^2+(dx_3-x_1dx_2)^2,\\
\nonumber\\
\label{Sl2} &&\mbox{SL}_2(\mathds{R}):\,\, d\tilde{s}^2=
\frac{1}{x_1^2}(dx_1^2+dx_2^2)+\Big(dx_3+\frac{dx_2}{x_1}\Big)^2,
\end{eqnarray}
\end{subequations}
which can be schematically written as
\begin{eqnarray}
d\tilde{s}^2=\sum_{I=1}^{3}{\omega}_I^2,
\end{eqnarray}
where the ${\omega}_I$ are the corresponding left-invariant
one-forms with $I=\{1,2,3\}$. For latter convenience, in what
follows, we will use the notation $I=(i,3)$ where $i$ ranges from
$1$ to $2$.

In order to be self-contained, we report the generalized Lifshitz
black holes solutions of standard five-dimensional General
Relativity, i. e. $G_{\mu\nu}+\Lambda g_{\mu\nu}=0$ found in
\cite{Cadeau:2000tj} and having horizon's topologies described by
the three-dimensional Solv and Nil geometries. In fact, these
solutions can be represented as follows
\begin{eqnarray}
ds^2=-r^{2z}f(r)dt^2+\frac{dr^2}{r^2f(r)}+\sum_{I=1}^{3}a_I\,r^{2q_I}{\omega}_I^2,
\label{CW}
\end{eqnarray}
where the $a_I$'s are constants that allow the introduction of an
eventual additional scale. In the case of the Solv black hole
solution, the set of parameters reads
\begin{eqnarray}
\label{SolvCW} &\mbox{Solv}:\qquad f(r)=1-\frac{M}{r^3},\quad\Big\{
z=1,\quad q_i=1, \quad q_3=0,\quad a_i=1,\quad
a_3=\frac{2}{3}\Big\},
\end{eqnarray}
while for black hole solution with Nil's horizon topology, the
parameters are given by
\begin{eqnarray}
\label{NilCW} &\mbox{Nil}:\quad f(r)=1-\frac{M}{r^{11/2}},\quad
\Big\{ z=3/2,\quad q_i=1,\quad q_3=2,\quad a_i=1,\quad
a_3=\frac{11}{2}\Big\}.
\end{eqnarray}
It is clear that both solutions satisfy asymptotically the
requirements given in (\ref{Lifshitznonst}-\ref{aniTransgener}).
More precisely, the Solv's solution (\ref{CW}-\ref{SolvCW}) is
asymptotically invariant under a one-parametric Lifshitz generalized
transformations (\ref{aniTransgener}) defined by
\begin{eqnarray}
t\to\lambda\,t,\qquad r\to\frac{1}{\lambda}\,r,\qquad x_1\to
\lambda^{1-\alpha}x_1,\qquad x_2\to \lambda^{1+\alpha}x_2,\qquad
x_3\to \lambda^{\alpha}x_3, \label{CWgenerTransfSolv}
\end{eqnarray}
while for the Nil's solution (\ref{CW}-\ref{NilCW}), one has two
anisotropic directions
\begin{eqnarray}
t\to\lambda^{\frac{3}{2}}\,t,\qquad r\to\frac{1}{\lambda}\,r,\qquad
x_i\to \lambda\, x_i,\qquad x_3\to \lambda^{2}\,x_3.
\label{CWgenerTransfNil}
\end{eqnarray}

In the present work, we propose to find the dyonic version of the
Solv (\ref{CW}-\ref{SolvCW}) and of the Nil's solution
(\ref{CW}-\ref{NilCW}). The interests for such study are multiple.
First of all, charged Lifshitz black holes are known to have rather
unconventional thermodynamical properties whose range is largely
spread from solutions with a Reissner-Nordstrom-like behavior
\cite{Tarrio:2011de} to zero-mass charged solutions
\cite{Pang:2009pd} including extremal solutions \cite{Liu:2014dva}.
The richness of these properties is essentially due to the
difficulty of "charging" the known Lifshitz solutions. This is in
contrast with the adS situation where an important class of charged
adS black hole solutions arise simply from the neutral
configurations turning on the Maxwell action. The situation is
radically different for the Lifshitz black holes where all the known
electrically charged Lifshitz black holes solutions of Einstein
gravity require, in addition to the Maxwell potential, some extra
fields materialized by scalar field with a dilatonic coupling
\cite{Tarrio:2011de} or a massive Proca field
\cite{Pang:2009pd,Alvarez:2014pra} or by considering nonminimal
coupling \cite{Babichev:2017rti}. In other words, the Maxwell field
alone is incompatible with the Lifshitz asymptotic for the
Einstein-Maxwell model. Nevertheless, this problem can be
circumvented in higher dimensions $D\geq 4$ where quadratic
corrections of the Einstein gravity can accommodate Maxwell charged
Lifshitz black holes \cite{Bravo-Gaete:2015xea}. The lesson learned
from these examples is that the presence of extra parameters in the
action permits to soften the incompatibility between the Maxwell
potential and the Lifshitz asymptotic. We would like to explore the
relevance of this observation in order to achieve our task of
charging the Solv and the Nil's solutions. More specifically, we
will consider a model described by the Einstein gravity with a
negative cosmological constant together with a scalar field and
various (at least $3$) $U(1)$ gauge fields with dilatonic-like
couplings. Indeed, dilatonic sources are usually good laboratories
for investigating charged black holes, see e. g. \cite{Dehghani,
Hendi:2015cra,Hendi:2017phi}. As shown below, the presence of more
than one $U(1)$ gauge field is mandatory in order to ensure the
Lifshitz asymptotic as well as the presence of the electric and
magnetic charges. In fact, the first gauge field guarantees the
Lifshitz asymptotic while the remaining Abelian fields sustain the
electric and magnetic charges\footnote{Note that for the purely
electrically Lifshitz charged black holes with planar, spherical or
hyperboloid horizon topology \cite{Tarrio:2011de}, two dilatonic
fields were at least required.}. We will also see that the dyonic
extensions of the of the Solv (\ref{CW}-\ref{SolvCW}) and of the
Nil's solution (\ref{CW}-\ref{NilCW}) with a multi-dilatonic source
present some interesting features. For example, the introduction of
the dilatonic source will extend the range of the dynamical
exponent. Indeed, while the vacuum Solv's (resp. Nil's) dyonic
solution requires $z=1$ (resp. $z=\frac{3}{2}$), their dyonic
extensions will exist for a Lifshitz dynamical exponent $z\geq 1$
(resp. $z\geq \frac{3}{2}$). Also, the dyonic solutions presented
below are quite different from those existing in the current
literature in the sense that their electric and magnetic charges are
proportional. This in turn implies that there does not exist a
purely electric or magnetic limit as it is the case for the
four-dimensional dyonic Reissner-Nordstrom solution.

The plan of the paper is organized as follows. In the next section,
we will explicitly present the five-dimensional model, its field
equations as well as the ansatz we will consider. In Secs. $3$ and
$4$, we will display the dyonic extensions of the Solv
(\ref{CW}-\ref{SolvCW}) and of the Nil's solution
(\ref{CW}-\ref{NilCW}). A detailed analysis of their thermodynamic
features will be provided showing that their electric and magnetic
charges are in fact proportional. In each case, we will check that
the electromagnetic version of the first law of the thermodynamics
is satisfied. In Sec. $5$, we will extend these results to the
so-called hyperscaling violating case with Solv and Nil's horizon
topologies. In this case, the Lagrangian model involves a Liouville
potential but without the cosmological constant. Interestingly
enough, for precise values of the hyperscaling violation factor the
electric contribution can be canceled yielding to purely
magnetically charged configurations. Finally, the last section is
dedicated to our conclusions.

%%%%%%%%%%%%%%%%%%%%%%%%%%%%%%%%%%%%%%%%%%%%%%%%%%%%%%%%%%%%%%%%%%%%%%
\section{Action, field equations and ansatz with Thurston geometries}
%%%%%%%%%%%%%%%%%%%%%%%%%%%%%%%%%%%%%%%%%%%%%%%%%%%%%%%%%%%%%%%%%%%%

As anticipated in the introduction, the five-dimensional action we
consider is given by the standard Einstein-Hilbert action with a
cosmological constant together with $N$ $U(1)$ gauge fields with
dilatonic-like couplings,
\begin{eqnarray}
S=\int d^5x\,\sqrt{-g}\left[\frac{R-2\Lambda}{2}-
\frac{1}{2}\partial_{\mu}\phi\partial^{\mu}\phi
-\frac{1}{4}\sum_{i=1}^{N}e^{\lambda_{i}\phi}F_{(i) \mu
\nu}F_{(i)}^{\mu \nu}\right], \label{action5d}
\end{eqnarray}
where as shown below $N=3$ in the case of the Solv's solution and
$N=4$ for the dyonic Nil's solution.

The equations of motions obtained by varying the action with respect
to the metric, the gauge vector fields and the scalar field
respectively  read
\begin{subequations}
\label{eqsm}
    \begin{eqnarray}
        && G_{\mu \nu}+\Lambda g_{\mu \nu}=T_{\mu\nu},\label{eqmotiong}\\
        &&\nabla_{\mu} \left(e^{\lambda_{i} \phi} F_{(i)}^{\mu
        \nu}\right)=0,\label{eqmotionAmu}\\
        &&\Box \phi= \sum_{i=1}^{N}\left(\frac{\lambda_{i}}{4} e^{\lambda_{i} \phi}
         F_{(i) \sigma \rho}
        F_{(i)}^{\sigma
        \rho}\right),\label{eqmotionphi}
    \end{eqnarray}
\end{subequations}
where the energy-momentum tensor $T_{\mu\nu}$ is defined as
\begin{eqnarray}
T_{\mu\nu}=&&\Big(\nabla_{\mu} \phi
        \nabla_{\nu} \phi-\frac{1}{2} g_{\mu \nu}
        \nabla_{\sigma}\phi \nabla^{\sigma}\phi\Big)+\sum_{i=1}^{N}
        \Big(e^{\lambda_{i} \phi}
        F_{(i)\mu \sigma} F_{(i) \nu}^{\phantom{\sigma \sigma} \sigma}
        -\frac{1}{4} g_{\mu \nu}
        e^{\lambda_{i}
        \phi} F_{(i) \sigma \rho}F_{(i)}^{\sigma
        \rho}\Big).
        \label{tmn}
\end{eqnarray}
In this paper, we will consider an ansatz for the metric of the form
(\ref{CW}), and in order for the metric ansatz (\ref{CW}) to be
asymptotically Lifshitz in the sense of
(\ref{Lifshitznonst}-\ref{aniTransgener}), we will require that
$\lim_{r\to\infty}f(r)=1$ and the left-invariant one-forms
${\omega}_I$ scale homogenously as ${\omega}_I\to
\lambda^{q_I}{\omega}_I$ under the dilatation transformations
$t\to\lambda^z \, t$ and $r\to\frac{1}{\lambda}r$.

%%%%%%%%%%%%%%%%%%%%%%%%%%%%%%%%%%%%%%%%%%%%%%%%%%%%%%%%%%%%%%%%%%%%%%%%%%%%%%
\section{Electromagnetic charged solution with a Solv's horizon topology\label{Solvsol}}
%%%%%%%%%%%%%%%%%%%%%%%%%%%%%%%%%%%%%%%%%%%%%%%%%%%%%%%%%%%%%%%%%%%%%%%%%%%%%%
We first report a charged dyonic black hole solution of the field
equations (\ref{eqsm}) for which the line element has a Solv's
horizon topology  parameterized as follows
\begin{eqnarray}
ds^2=&& -r^{2z}\Big[1-m\left(\frac{r_h}{r}\right)^{z+2}
+(m-1)\left(\frac{r_h}{r}\right)^{2z+2}\Big]dt^2+
\frac{dr^2}{r^{2}\left[1-m\left(\frac{r_h}{r}\right)^{z+2}+(m-1)\left(\frac{r_h}{r}\right)^{2z+2}\right]}
 \nonumber\\
&&\hspace{1in} +r^{2} x_3^{2}\,dx_1^2+r^{2}\,\frac{dx_2^2}{x_3^{2}}
+\left(\frac{2}{z+2}\right)\,\frac{dx_3^2}{x_3^{2}}.\label{sol1asolvdyon}
\end{eqnarray}
The matter fields associated to this spacetime metric read
\begin{eqnarray}
\label{fieldSolv}
 &&e^{\phi}=r^{\sqrt{{2 (z-1)}}},\qquad
F_{(1)rt}=\sqrt
{(z+2)(z-1)}\,{r}^{z+1},\\
&& F_{(2)rt}=\sqrt{z (m-1)}\left(\frac{r_h}{r}\right)^{z+1},\qquad
F_{(3)x_1 x_2}=\sqrt {z(m-1)}\, {{r_h}^{z+1}},\nonumber
\end{eqnarray}
and the solution exists provided that the coupling constants are
chosen as
\begin{eqnarray}
\label{lambdasolvdyn} \Lambda=-\frac{(z+2)^{2}}{2},\qquad
\lambda_{1}=-\frac{4}{\sqrt{2(z-1)}},\qquad
\lambda_2=\sqrt{2(z-1)},\qquad \lambda_3=-\sqrt{2(z-1)}.
\end{eqnarray}
Before providing a complete thermodynamics analysis of the Solv's
dyonic solution, some additional comments are needed. Firstly, the
existence of the Solv's solution is ensured for at least three
Abelian gauge fields with dilatonic-like couplings. Secondly, we
find judicious to parameterize the metric solution as in
(\ref{sol1asolvdyon}) which makes clear that $r_h$ stands for the
location of the event horizon. Nevertheless, as shown below, the two
integration constants $m$ and $r_h$ will be identified with the
mass, the electric and magnetic charges. In addition, the metric
function appearing in (\ref{sol1asolvdyon}) will have a
Reissner-Nordstrom-like form. It will also become clear after the
thermodynamics analysis that the electric and magnetic charges are
proportional which in turn explains the mismatch between the number
of integration constants and of charges. The range of the Lifshitz
dynamical exponent is given by $z\geq 1$. In fact, even if the
coupling constant $\lambda_1$ as defined in (\ref{lambdasolvdyn})
blows up in the limit $z=1$, the first dilaton Lagrangian in the
action $e^{\lambda_1\phi}F_{(1)\mu\nu}F_{(1)}^{\mu\nu}\to 0$ as
$z\to 1$. More precisely, the limiting adS case $z=1$ reduces to the
dyonic solution recently found in \cite{Arias:2017yqj} in the
absence of the scalar field or to the vacuum solution for $m=1$
(\ref{CW}-\ref{SolvCW}), see Ref. \cite{Cadeau:2000tj}. Hence,
interestingly enough, the fact of turning on the dilatonic source
permits to extend the range of the dynamical exponent to be $z\geq
1$. Consequently, the asymptotic metric is invariant under the
following one-parametric Lifshitz generalized dilatation
transformations (\ref{aniTransgener}) extending those of the vacuum
sector (\ref{CWgenerTransfSolv}) and defined by
$$
t\to\lambda^z\,t,\qquad r\to\frac{1}{\lambda}\,r,\qquad x_1\to
\lambda^{1-\alpha}x_1,\qquad x_2\to \lambda^{1+\alpha}x_2,\qquad
x_3\to \lambda^{\alpha}x_3.
$$
Also, the scalar field is defined up to a constant $c$, and this
constant can be put to zero without any loss of generality since it
is a symmetry of the dilaton action represented as $\phi\to \phi-c$
and $A_{(i)}\to e^{\frac{\lambda_i c}{2}}A_{(i)}$. Finally, we would
like  to point out  an interesting fact concerning the
electro-magnetic duality. In the hyperplane defined by
$x_3=\mbox{cst}$, the electric and magnetic fields are dual in the
sense that
\begin{eqnarray}
\star\, F_{(2)}=F_{(3)}, \label{Hodgeduality}
\end{eqnarray}
where the Hodge dual operator $\star$ is defined for the
four-dimensional metric defined by $x_3=\mbox{cst}$.

We now turn to the thermodynamics study of the Solv's charged dyonic
solution (\ref{sol1asolvdyon}-\ref{lambdasolvdyn}). As was shown in
\cite{Gibbons:1976ue}, the partition function for a thermodynamics
ensemble may be identified with the Euclidean path integral in the
saddle point approximation around the Euclidean continuation of the
solution. In the present case, we will deal with a reduced action
principle with a static Euclidean metric endowed by a Solv's horizon
topology. More precisely, the Euclidean ansatz for this mini
superspace configuration is given by the following line element
\begin{eqnarray}
ds^2= N^2(r) F(r)d\tau^2+\frac{dr^2}{F(r)}+r^{2}
x_3^{2}\,dx_1^2+r^{2}\,\frac{dx_2^2}{x_3^{2}}
+\left(\frac{2}{z+2}\right)\,\frac{dx_3^2}{x_3^{2}},
\end{eqnarray}
with matter fields given as
\begin{eqnarray*}
A_{(i)\mu}dx^{\mu}=A_{(i)\tau}(r)d\tau,\quad
A_{(3)\mu}dx^{\mu}=A_{(3)x_1}(x_2)dx_1+A_{(3)x_2}(x_1)dx_2,\quad
\phi=\phi(r),
\end{eqnarray*}
with $i=1,2$. In the Euclidean continuation, the range of the radial
coordinate is from the horizon $r_h$ to infinity and the Euclidean
time $\tau=it$ is compactified as $\tau\in [0,\beta]$ where $\beta$
stands for the inverse of the temperature $\beta=T^{-1}$. As usual,
using this ansatz, a reduced action can be written in a "Hamilton
form". Nevertheless, because of the presence of the magnetic field
$F_{(3)}$, we will carefully derive this Hamilton form in various
steps.

The Euclidean action denoted $I_E$ for the previous ansatz is
schematically decomposed in five pieces as
\begin{eqnarray}
I_E=I_{\tiny{\mbox{EH}}}+I_{\tiny{\mbox{kin}}}+\sum_{i=1}^2I_{F_{(i)}}+I_{F_{(3)}}+B_E.
\label{IEgeneral}
\end{eqnarray}
The first two terms correspond to the Einstein-Hilbert piece and the
kinetic term of the scalar field, the $I_{F_{(i)}}$'s are the
electric dilatonic parts of the action while $I_{F_{(3)}}$ stands
for the magnetic part(s) and $B_E$ is a boundary term fixed in such
a way that the reduced action $I_E$ has a well-defined extremum,
that is $\delta I_E=0$. As shown below, the Euclidean action
on-shell reduces to the boundary term and is related to the Gibbs
free energy ${\cal G}$ as
\begin{eqnarray}
I_E=\beta {\cal G}=\beta\left({\cal M}-\Phi_e{\cal Q}_e-\Phi_m{\cal
Q}_m\right)-{\cal S}, \label{Gibbs}
\end{eqnarray}
where ${\cal M}$ is the mass, ${\cal S}$ the entropy, $\Phi_e$
(resp. $\Phi_m$) corresponds to the electric (resp. magnetic)
potential and ${\cal Q}_e$ (resp. ${\cal Q}_m$) represents the
electric (resp. magnetic) charge. Note that we opt for the formalism
of the grand canonical ensemble where the temperature as well the
electric and magnetic potentials are fixed.

For the dyonic Solv's solution, we found that the different pieces
of the reduced Euclidean action are given by
\begin{eqnarray*}
&&I_{\tiny{\mbox{EH}}}=\vert \Omega_{\tiny{\mbox{Solv}}}\vert
\frac{\beta}{2}\sqrt{\frac{2}{z+2}}\int
_{r_h}^{\infty}N(r)\left[2\Lambda
r^2+r^2z+2rF(r)^{\prime}+2r^2+2F(r)\right]dr,\\
&&I_{\tiny{\mbox{kin}}}=\vert \Omega_{\tiny{\mbox{Solv}}}\vert
\frac{\beta}{2} \sqrt{\frac{2}{z+2}}\int
_{r_h}^{\infty}N(r)\left[r^2 F(r) \,(\phi(r)^{\prime})^2\right]dr,\\
&& I_{F_{(i)}}=\beta \int \Big[A_{(i)\tau}\partial_r {\cal
P}_{(i)}(r,x_i,x_3)+\frac{N(r)}{2r^2}x_3\sqrt{\frac{z+2}{2}}e^{-\lambda_i\phi}\,{\cal
P}_{(i)}(r,x_i,x_3)^2\Big]dx_1dx_2dx_3dr,\\
&&I_{F_{(3)}}=\frac{\beta}{2}\sqrt{\frac{2}{z+2}}\int
\frac{N(r)}{r^2x_3}e^{\lambda_3\phi}\,\left(\partial_{x_1}A_{(3)x_2}-\partial_{x_2}A_{(3)x_1}\right)^2\,dx_1dx_2dx_3dr.
\end{eqnarray*}
In these expressions, we have defined $\vert
\Omega_{\tiny{\mbox{Solv}}}\vert$ to be the volume element of the
compact Solv's spacetime (\ref{Solvg}), that is
\begin{eqnarray}
\vert \Omega_{\tiny{\mbox{Solv}}}\vert=\int_{\Omega_1\times
\Omega_2\times \Omega_3}
dx_1\,dx_2\,\frac{dx_3}{x_3}=\vert\Omega_1\vert\,
\vert\Omega_2\vert\, \ln\vert \Omega_3\vert, \label{solvvolume}
\end{eqnarray}
where the $\Omega_I$'s  for $I=1,2, 3$ stand for the compact ranges
of the horizon coordinates $x_I$. On the other hand, the ${\cal
P}_{(i)}$'s are the conjugate momenta of the electric potential
fields $A_{(i)}$ for $i=1,2$. We can note that the last two
integrals $I_{F_{(i)}}$ and $I_{F_{(3)}}$ still involve the
four-dimensional volume element; this is due to the fact that the
electric conjugate momentum and the magnetic gauge field are allowed
to depend on the horizon coordinates $x_i$ and $x_3$. Nevertheless,
this dependence can be specified through the field equations
associated to the Euclidean action. Indeed, the variation of $I_E$
with respect to the conjugate momenta ${\cal P}_{(i)}$ implies that
the conjugate momenta are separable in the following way
$$
{\cal P}_{(i)}=\frac{{\cal
\bar{P}}_{(i)}(r)}{x_3}\qquad\mbox{with}\quad  {\cal
\bar{P}}_{(i)}(r)
=r^2\sqrt{\frac{2}{z+2}}\frac{e^{\lambda_i\phi}}{N}\partial_r
A_{(i)\tau},\qquad \mbox{for}\quad i=1,2.
$$
On the other hand, the variation with respect to the magnetic gauge
field $A_{(3)}$ forces the magnetic gauge field to be lineal in
$x_i$. Hence, under these last considerations, the reduced action
$I_E$ can be written in a Hamilton form as {\small \begin{eqnarray}
\label{redSolvaction} && I_E=\beta \vert
\Omega_{\tiny{\mbox{Solv}}}\vert \int _{r_h}^{\infty} \left(N{\cal
H}+\sum_{i=1}^2A_{(i)\tau}{\cal
\bar{P}}_{(i)}^{\prime} \right)dr+B_E,\\
&& {\cal H}=\sqrt{\frac{2}{z+2}}\left[\Lambda
r^2+r^2\Big(1+\frac{z}{2}\Big)+F+rF^{\prime}+\frac{r^2}{2} F
(\phi^{\prime})^2+\sum_{i=1}^2 \frac{z+2}{4r^2}
e^{-\lambda_i\phi}{\cal
\bar{P}}_{(i)}^2+\frac{e^{\lambda_3\phi}}{2r^2}(F_{(3)x_1x_2})^2\right]\nonumber.
\end{eqnarray}}
It is reassuring to check that the field equations obtained by
varying the reduced action $I_E$ (\ref{redSolvaction}) with respect
to $N, F, \phi, \bar{{\cal P}}_{(i)}, A_{(i)}$ and $A_{(3)}$ are
consistent with the original equations of motion (\ref{eqsm}).

Now, we are in position to determine the boundary term $B_E$ that
will encode all the thermodynamics features of the solution. This
term is fixed by requiring that the total action has an extremum
$\delta I_E=0$ with
$$
\delta I_E=\beta \vert \Omega_{\tiny{\mbox{Solv}}}\vert
\left[\sqrt{\frac{2}{z+2}} N(r\delta
F+r^2F\phi^{\prime}\delta\phi)+\sum_{i=1}^2A_{(i)\tau}\delta{\cal
\bar{P}}_{(i)}\right]_{r_h}^{\infty}+\delta I_{F_{(3)}}+\delta B_E.
$$
The variation of the magnetic part $\delta I_{F_{(3)}}$ must be done
with care,
\begin{eqnarray*}
\delta I_{F_{(3)}}&=&\beta\sqrt{\frac{2}{z+2}}\ln\vert\Omega_3\vert
\int
dr \frac{N e^{\lambda_3\phi}}{r^2}\epsilon^{ij} \left[\partial_{x_i}A_{(3)x_j}\delta A_{(3)x_j}\vert\Omega_j\vert \right]_{x_i\in\Omega_i}\\
\\
&=&\beta\vert \Omega_{\tiny{\mbox{Solv}}}\vert
\sqrt{\frac{2}{z+2}}\sqrt{z(m-1)}\,r_h^{z+1}\delta\left(\sqrt{z(m-1)}\,r_h^{z+1}\right)\int_{r_h}^{\infty}
dr \frac{N e^{\lambda_3\phi}}{r^2},
\end{eqnarray*}
where $\epsilon^{ij}$ is the totally antisymmetric tensor with
$\epsilon^{12}=1$. Note that in the second line, we have used the
fact that $A_{(3)}$ is linear in $x_i$ (\ref{fieldSolv}), and this
also explains the reason for which the volume element of the Solv's
geometry (\ref{solvvolume}) appears. This variation of the magnetic
piece is analogous  to what occurs in the magnetically charged
Reissner-Nordstrom solution or to what have been done recently in
the case of adS$_4$ dyonic black holes \cite{Cardenas:2016uzx}. For
the Solv's solution (\ref{sol1asolvdyon}-\ref{lambdasolvdyn}) with
metric functions $N$ and $F$ identified as
$$
N(r)=r^{z-1},\qquad
F(r)=r^{2}\left[1-m\left(\frac{r_h}{r}\right)^{z+2}+(m-1)\left(\frac{r_h}{r}\right)^{2z+2}\right],
$$
a straightforward computation permits to obtain the variation of the
boundary term {\small \begin{eqnarray*} \delta B_E=\beta \vert
\Omega_{\tiny{\mbox{Solv}}}\vert\Bigg[
\sqrt{\frac{2}{z+2}}\,\left(\delta(mr_h^{z+2})-
\frac{2\pi}{\beta}\delta
r_h^2\right)-\left[A_{(2)\tau}(\infty)-A_{(2)\tau}(r_h)\right]\,\delta\left(\sqrt{\frac{2z(m-1)}{z+2}}\,r_h^{z+1}\right)\\-
\sqrt{\frac{2(m-1)}{z+2}}\,r_h\,\delta\left(\sqrt{m-1}\,r_h^{z+1}\right)\Bigg],
\end{eqnarray*}}
where, as usual, in order to avoid conical singularity, we require
that $N\delta F\vert_{r_h}=-\frac{4\pi}{\beta}\delta r_h$. We also
identify the electric potential $\Phi_e$ as the difference of the
gauge field between the infinity and the event radius, i. e.
$$
\Phi_e=A_{(2)\tau}(\infty)-A_{(2)\tau}(r_h)=\sqrt{\frac{m-1}{z}}\,r_h,
$$
and hence the second piece in $\delta B_E$  is identified with the
electric variation $-\Phi_e\delta {\cal Q}_e$. Analogously, the last
variation in $\delta B_E$ must correspond to the magnetic variation
$-\Phi_m\delta {\cal Q}_m$, see (\ref{Gibbs}). Nevertheless,
contrary to the electric part, there is a priori no way of
identifying the magnetic potential, and hence the magnetic potential
and charge can only be determined up to two constants, namely
$$
\Phi_m=A\,\sqrt{m-1}\,\,r_h,\qquad {\cal Q}_m=B\,\sqrt{m-1}\,\vert
\Omega_{\tiny{\mbox{Solv}}}\vert\,r_h^{z+1},\qquad
AB=\sqrt{\frac{2}{z+1}}.
$$
We can note from now that independently of the fact that the
constants $A$ and $B$ are not fixed, it is clear that the electric
and magnetic charge are proportional (see below for the electric
charge). However, since the electric and magnetic variations are
equal $-\Phi_e\delta {\cal Q}_e=-\Phi_m\delta {\cal Q}_m$, and
because of the electromagnetic duality in the hyperplane defined by
$x_3=\mbox{cst}$ (\ref{Hodgeduality}), one can suppose that
$\Phi_e=\Phi_m$ and ${\cal Q}_e={\cal Q}_m$. Finally, in the
formalism of the grand canonical ensemble the boundary term can be
expressed as
\begin{eqnarray*}
B_E=\beta \vert \Omega_{\tiny{\mbox{Solv}}}\vert\Bigg[
\sqrt{\frac{2}{z+2}}\,mr_h^{z+1}-(\Phi_e+\Phi_m)\left(\sqrt{\frac{2z(m-1)}{z+2}}\,r_h^{z+1}\right)\Bigg]-
2\pi\sqrt{\frac{2}{z+2}}\vert \Omega_{\tiny{\mbox{Solv}}}\vert
r_h^2.
\end{eqnarray*}
Since the on-shell Euclidean action reduces to the boundary term
$I_{E\vert_{\mbox{\tiny{on-shell}}}}=B_E$, the different
thermodynamics quantities can easily be determined through
(\ref{Gibbs}) yielding to
\begin{eqnarray}
&&{\cal M}=\vert \Omega_{\tiny{\mbox{Solv}}}\vert
\sqrt{\frac{2}{z+2}}\,m\, r_h^{z+2},\qquad
T=\frac{\left(2z+2-mz\right) r_h^z}{4\pi},\qquad
{\cal S}=2\pi \vert \Omega_{\tiny{\mbox{Solv}}}\vert \sqrt{\frac{2}{z+2}}\,r_h^2,\nonumber\\
\\
&&\Phi_e=\Phi_m=\sqrt{\frac{m-1}{z}}\,r_h,\qquad {\cal Q}_e={\cal
Q}_m=\vert \Omega_{\tiny{\mbox{Solv}}}\vert
\sqrt{\frac{2z(m-1)}{z+2}}\,r_h^{z+1}.
\nonumber\label{thermoqtesSolv}
\end{eqnarray}
It is straightforward to check the validity of the first law of
thermodynamics
\begin{eqnarray}
d{\cal M}=Td{\cal S}+\Phi_e d{\cal Q}_e+\Phi_m d{\cal Q}_m.
\label{1stlawSolv}
\end{eqnarray}
Just to conclude this section, we note that the metric solution
(\ref{sol1asolvdyon}) can now be re-written
 in the Reissner-Nordstrom-like form as
\begin{eqnarray}
ds^2=&& -r^{2z}f(r)dt^2+ \frac{dr^2}{r^{2}f(r)}+r^{2}
x_3^{2}\,dx_1^2+r^{2}\,\frac{dx_2^2}{x_3^{2}}
+\left(\frac{2}{z+2}\right)\,\frac{dx_3^2}{x_3^{2}},\nonumber\\
\\
f(r)=&&1-\sqrt{\frac{z+2}{2}}\frac{{\cal M}}{\vert
\Omega_{\tiny{\mbox{Solv}}}\vert\, r^{z+2}} +\frac{(z+2)}{4z \vert
\Omega_{\tiny{\mbox{Solv}}}\vert^2}\frac{{\cal Q}_e^2+{\cal
Q}_m^2}{r^{2(z+1)}}.\nonumber \label{sol1asolvdyonRN}
\end{eqnarray}
From this last expression, one observes that in the adS limit $z=1$,
even if the fall off  of the mass term is more faster than in the
standard five-dimensional Reissner-Nordstrom case, one still get a
finite and nonzero value of the mass for the Solv's dyonic solution.

%%%%%%%%%%%%%%%%%%%%%%%%%%%%%%%%%%%%%%%%%%%%%%%%%%%%%%%%%%%%%%%%%%%%%%%%
\section{Purely Lifshitz dyonic solution with a Nil's horizon topology}
%%%%%%%%%%%%%%%%%%%%%%%%%%%%%%%%%%%%%%%%%%%%%%%%%%%%%%%%%%%%%%%%%%%%%%%%%
We now present the dyonic extension of the Nil's solution
(\ref{CW}-\ref{NilCW}). In this case, it is possible to find the
following class of solution with a line element that reads
\begin{eqnarray}
ds^2&=& -r^{2z}\left[1-m\left(\frac{r_h}{r}\right)^{z+4}
+(m-1)\left(\frac{r_h}{r}\right)^{2z+4}\right]dt^2+
\frac{dr^2}{r^{2}\left[1-m\left(\frac{r_h}{r}\right)^{z+4}+
(m-1)\left(\frac{r_h}{r}\right)^{2z+4}\right]}\nonumber\\
 &&\hspace{.8in}+r^{2}dx_1^2+r^{2}
dx_2^2 +\left(z+4\right) r^{4} (dx_3-x_1dx_2)^2.\label{sol2anildyon}
\end{eqnarray}
The gauge and scalar fields associated to this line element are
given by
\begin{eqnarray}
\label{fieldsNil} &&e^{\phi}=r^{\sqrt{2(2z-3)}},\qquad \qquad F_{(1)
rt}=\frac{\sqrt {2 \,(2\,z-3)(z+4)}}{2}\,{{r}^{z+3}},\\
&& F_{(2)rt}=\frac{\sqrt {2\, z (m-1) }r_h^{z+2}}
 {r^{z+1}},\quad
F_{(3)x_1 x_2}=-x_1\sqrt{ z \left( m-1 \right) \left( z+4
\right)}\,r_h^{z+2},\nonumber\\
\nonumber\\
 &&  F_{(3)x_1 x_3}=\sqrt{ z \left( m-1 \right) \left(
z+4 \right)}\,r_h^{z+2},\quad F_{(4)x_2 x_3}=\sqrt{ z \left( m-1
\right) \left( z+4 \right)} r_h^{z+2},\nonumber
\end{eqnarray}
and the parameters must be chosen as
\begin{eqnarray*}
\Lambda&=&-{\frac { \left( z+4 \right)  \left( z+3 \right)
}{2}},\qquad \lambda_{1}=-{\frac {8}{\sqrt {2 (2\,z-3)}}},\qquad
\lambda_2={\frac {2\left( z-2\right) }{\sqrt
{2 (2\,z-3)}}},\nonumber\\
\lambda_3&=&\lambda_4=-{\frac { 2\left(z-1\right)}{\sqrt {2
(2\,z-3)}}}. \label{lambdanildyon}
\end{eqnarray*}
Few comments can be made concerning this dyonic solution with Nil's
horizon. Firstly, the Nil dyonic solution requires at least four
$U(1)$ gauge fields with dilatonic couplings and is valid for a
dynamical exponent $z\geq \frac{3}{2}$. As before, the limiting case
$z=\frac{3}{2}$ with $m=1$ (that is without electromagnetic charges)
reduces to the vacuum Nil's solution (\ref{CW}-\ref{NilCW}).
Secondly, the solution does not exhibit a such electromagnetic
duality (\ref{Hodgeduality}) as was for the Solv's solution. In
addition, the dyonic Nil's solution can be likened to a Lifshitz
black hole whose asymptotic symmetries contain a generalized
dilatation transformation with two anisotropic direction
(\ref{aniTransgener}) given by
$$
t\to\lambda^z\,t,\qquad r\to\frac{1}{\lambda}\,r,\qquad x_i\to
\lambda\, x_i,\qquad x_3\to \lambda^{2}\,x_3.
$$
Finally, as in the previous case, the two integration constants will
be shown to represent the mass and the electric/magnetic charges.

Let us now study the thermodynamics properties of the Nil's
solution. Following the same lines as those presented in details for
the Solv's solution, we consider the following Euclidean ansatz
\begin{eqnarray*}
&&ds^2= N^2(r) F(r)d\tau^2+\frac{dr^2}{F(r)}+r^{2}dx_1^2+r^{2}
dx_2^2 +\left(z+4\right) r^{4} (dx_3-x_1dx_2)^2,\\
\\
&&A_{(i)\mu}dx^{\mu}=A_{(i)\tau}(r)d\tau,\quad
A_{(3)\mu}dx^{\mu}=A_{(3)I}(x_J)dx^I,\quad
A_{(4)\mu}dx^{\mu}=A_{(4)I}(x_J)dx^I,\quad \phi=\phi(r).
\end{eqnarray*}

For this class of ansatz, the  Euclidean action (\ref{IEgeneral}) is
decomposed as
\begin{eqnarray*}
&&I_{\tiny{\mbox{EH}}}=\vert
\Omega_{\tiny{\mbox{Nil}}}\vert\beta\sqrt{z+4}\int
_{r_h}^{\infty}N(r)\left[7r^2F+r^4\left(\frac{z}{4}+\Lambda+1\right)+2r^3
F^{\prime}\right]dr,\\
&&I_{\tiny{\mbox{kin}}}=\vert \Omega_{\tiny{\mbox{Nil}}}\vert
\frac{\beta}{2} \sqrt{\frac{2}{z+2}}\int
_{r_h}^{\infty}N(r)\left[r^4 F(r) \,(\phi(r)^{\prime})^2\right]dr,\\
&& I_{F_{(i)}}=\vert \Omega_{\tiny{\mbox{Nil}}}\vert \beta \int
\Big[A_{(i)\tau}\partial_r {\cal
\bar{P}}_{(i)}+\frac{N}{r^4\sqrt{z+4}}e^{-\lambda_i\phi}\,{\cal
\bar{P}}_{(i)}^2\Big]\,dr,
\end{eqnarray*}
and the magnetic pieces read
\begin{eqnarray*}
I_{F_{(3,4)}}=\frac{\beta}{2}\sqrt{z+4}\int \Bigg\{&& N
e^{\lambda_3\phi}\,\left[F_{(3)x_1x_2}^2+2x_1F_{(3)x_1x_2}F_{(3)x_1x_3}+\left(\frac{1}{
r^2(z+4)}+x_1^2\right)F_{(3)x_1x_3}^2\right]\\
&&+ N e^{\lambda_4\phi}\frac{F_{(4)x_2x_3}^2}{r^2(z+4)}\Bigg\}dx_1
dx_2 dx_3 dr.
\end{eqnarray*}
In these expressions, we have defined $\vert
\Omega_{\tiny{\mbox{Nil}}}\vert$ to be the volume element of the Nil
geometry (\ref{Nilg})
\begin{eqnarray}
\vert \Omega_{\tiny{\mbox{Nil}}}\vert=\int_{\Omega_1\times
\Omega_2\times \Omega_3} dx_1\,dx_2\,dx_3=\vert\Omega_1\vert\,
\vert\Omega_2\vert\, \vert \Omega_3\vert. \label{nilvolume}
\end{eqnarray}
The dependence of the magnetic field strengths $F_{(3)}$ and
$F_{(4)}$ on the Thurston's coordinates $x_I$ can be fixed by
varying the total action (\ref{IEgeneral}) w. r. t. $A_{(3)I}$ and
$A_{(4)I}$. In doing so, one obtains that
$$
F_{(3)x_1x_3}=-\frac{1}{x_1}F_{(3)x_1x_2}=\mbox{cst},\qquad
F_{(4)x_2x_3}=\mbox{cst}.
$$
Using this last result, the $x_I-$dependence of the magnetic action
$I_{F_{(3,4)}}$ is canceled out, and hence the reduced Euclidean
action (\ref{IEgeneral}) can be written in Hamilton form depending
only on the radial coordinate as
\begin{eqnarray*}
I_E=\beta \vert \Omega_{\tiny{\mbox{Nil}}}\vert \int _{r_h}^{\infty}
\left(N{\cal H}+\sum_{i=1}^2A_{(i)\tau}{\cal \bar{P}}_{(i)}^{\prime}
\right)dr+B_E,
\end{eqnarray*}
where the Hamiltonian reads
\begin{eqnarray}
{\cal
H}={\sqrt{z+4}}\Bigg[&&7r^2F+r^4\left(\frac{z}{4}+\Lambda+1\right)+2r^3
F^{\prime}+\frac{r^4}{2} F (\phi^{\prime})^2+\sum_{i=1}^2
\frac{e^{-\lambda_i\phi}}{(z+4)r^4}{\cal
\bar{P}}_{(i)}^2+\frac{e^{\lambda_3\phi}}{2r^{2}(z+4)}F_{(3)x_1x_3}^2\nonumber\\
&&
+\frac{e^{\lambda_4\phi}}{2r^{2}(z+4)}F_{(4)x_2x_3}^2\Bigg]\nonumber,
\end{eqnarray}
As explained before, the boundary term encodes all the
thermodynamical features of the solution. The boundary is fixed by
requiring that the on-shell Euclidean action has a well-defined
extremum $\delta I_E=0$ with
$$
\delta I_E=\beta \vert \Omega_{\tiny{\mbox{Nil}}}\vert
\left[\sqrt{z+4} N(2r^3\delta
F+r^4F\phi^{\prime}\delta\phi)+\sum_{i=1}^2A_{(i)\tau}\delta{\cal
\bar{P}}_{(i)}\right]_{r_h}^{\infty}+\delta I_{F_{(3,4)}}+\delta
B_E,
$$
where $\delta I_{F_{(3,4)}}$ stands for the variation of the
magnetic dilaton parts of the action
\begin{eqnarray*}
\delta I_{F_{(3,4)}}=\frac{\beta}{\sqrt{z+4}}\Bigg\{\int dr\,
\frac{N e^{\lambda_3\phi}}{r^2}\,\sigma_{(3)}^{ij}\Bigg[\partial_i
A_{(3)j}\,\delta A_{(3)j}\,\vert\Omega_2\vert \vert \Omega_j\vert
\Bigg]_{x_i\in\Omega_i}\\
+\int dr\, \frac{N
e^{\lambda_4\phi}}{r^2}\,\sigma_{(4)}^{ij}\Bigg[\partial_i
A_{(4)j}\,\delta A_{(4)j}\,\vert\Omega_1\vert \vert \Omega_j\vert
\Bigg]_{x_i\in\Omega_i}\Bigg\},
\end{eqnarray*}
where the non-vanishing components of  $\sigma_{(3)}^{ij}$ and
$\sigma_{(4)}^{ij}$ are given by
$\sigma_{(3)}^{13}=-\sigma_{(3)}^{31}=1$ and
$\sigma_{(4)}^{23}=-\sigma_{(4)}^{32}=1$.

After some computations, for the Nil's solution with metric
functions and conjugate momenta given by
$$
N(r)=r^{z-1},\quad
F(r)=r^{2}\left[1-m\left(\frac{r_h}{r}\right)^{z+4}+
(m-1)\left(\frac{r_h}{r}\right)^{2z+6}\right],\quad {\cal
\bar{P}}_{(i)}=\frac{\sqrt{z+4}\,r^4}{N(r)}e^{\lambda_i\phi}\partial_r
A_{(i)\tau},
$$
one obtains for the variation of the boundary term
\begin{eqnarray*}
\delta B_E=\beta \vert \Omega_{\tiny{\mbox{Nil}}}\vert\Bigg[
\sqrt{z+4}\,\left(\delta(2m\,r_h^{z+4})- \frac{2\pi}{\beta}\delta
r_h^4\right)-\Phi_e\,\delta\left(\sqrt{2z(z+4)(m-1)}
r_h^{z+2}\right)\\-2
\sqrt{(m-1)}\,r_h^2\,\delta\left(\sqrt{(m-1)(z+4)}\,r_h^{z+2}\right)\Bigg],
\end{eqnarray*}
where the electric potential is given by
$\Phi_e=\sqrt{\frac{2(m-1)}{z}} r_h^2$. Finally, the boundary term
in the formalism of the grand canonical ensemble is expressed as
$$
B_E=\beta \vert \Omega_{\tiny{\mbox{Nil}}}\vert\Bigg[2\sqrt{z+4}
m\,r_h^{z+4}-\Phi_e{\cal Q}_e-\Phi_m{\cal
Q}_m\Bigg]-2\pi\sqrt{z+4}\vert \Omega_{\tiny{\mbox{Nil}}}\vert.
$$
However, in this case, we can not use a duality argument in order to
properly fix the magnetic potential. Hence, the magnetic potential
and charge will only be defined up to two constants $A$ and $B$, and
the thermodynamics quantities  read off from (\ref{Gibbs}) are given
by
\begin{eqnarray}
&&{\cal M}=2\vert \Omega_{\tiny{\mbox{Nil}}}\vert \sqrt{z+4}\,m\,
r_h^{z+4},\qquad T=\frac{(2z-mz+4) r_h^z}{4\pi},\qquad
{\cal S}=2\pi \vert \Omega_{\tiny{\mbox{Nil}}}\vert \sqrt{z+4}\,r_h^4,\nonumber\\
\nonumber\\
 &&\Phi_e=\sqrt{\frac{2(m-1)}{z}} r_h^2,\qquad {\cal
Q}_e=\vert \Omega_{\tiny{\mbox{Nil}}}\vert
\sqrt{2z(z+4)(m-1)}\,r_h^{z+2},\\
\nonumber\\
 &&\Phi_m=A\sqrt{m-1}\,r_h^2,\qquad {\cal Q}_m=\vert
\Omega_{\tiny{\mbox{Nil}}}\vert\,B\sqrt{m-1}\,r_h^{z+2},\qquad
AB=2\sqrt{z+4}. \nonumber \label{thermoqtesNil}
\end{eqnarray}
Nevertheless, in spite of the "arbitrariness" concerning the
magnetic potential and charge, the first law of thermodynamics only
requires the product of the constants $A$ and $B$, and it is a
matter of check to see that the first law (\ref{1stlawSolv})
effectively holds, and just need that $AB=2\sqrt{z+4}$.

%%%%%%%%%%%%%%%%%%%%%%%%%%%%%%%%%%%%%%%%%%%%%%%%%%%%%%%%%%%%%%%%%%%
\section{Dyonic solutions with a hyperscaling violation factor}
%%%%%%%%%%%%%%%%%%%%%%%%%%%%%%%%%%%%%%%%%%%%%%%%%%%%%%%%%%%%%%%%%%%
One of the main interest in extending the adS/CFT correspondence to
other areas of the physics was precisely to have a better
understanding of strongly coupled systems of the condensed matter
physics. In condensed matter physics, the notion of quantum phase
transition is of great importance and it occurs at some critical
point where the system may display a hyperscaling violation
reflected by the fact that the entropy does not scale with its
spatial dimensionality. From the gravity side, such hyperscaling
violating systems can be described by the so-called hyperscaling
violating metrics \cite{Charmousis:2012dw} which are conformally
related to the Lifshitz metric as
\begin{eqnarray}
ds^2=\frac{1}{r^{\frac{2\theta}{D-2}}}\left[-r^{2z}dt^2+\frac{dr^2}{r^2}+r^2\sum_{i=1}^{D-2}dx_i^2\right],
\label{hvm}
\end{eqnarray}
in such a way that the anisotropic transformations (\ref{aniTransf})
act now as a conformal transformation, i. e. $ds^2\to
\lambda^{\frac{2\theta}{D-2}} ds^2$. Here the parameter $\theta$ is
the so-called the {\it hyperscaling violation factor} responsible of
the violation of the hyperscaling property. Of course, hyperscaling
violation black holes refer to black hole solutions whose asymptotic
forms match with the metric (\ref{hvm}), see e. g.
\cite{Dong:2012se,Alishahiha:2012qu,HVMBH1,HVMBH2}. As before, this
notion of hyperscaling violation black holes  can be enlarged by
relaxing the fact that the topology of the horizon is flat but still
requiring that the generalized dilatation transformations
(\ref{aniTransgener}) act as a conformal transformation for the
metric solution in the asymptotic region. In this context, a
hyperscaling violation black hole of General Relativity was found in
\cite{Hassaine:2015ifa} where the horizon topology is modeled by the
Nil's geometry and where the dynamical exponents are $z=\frac{3}{2}$
and $\theta=\frac{9}{2}$. As done previously, we will see that this
vacuum solution can be electromagnetically charged by turning on a
dilatonic source. For this purpose, we consider a slightly different
action than (\ref{action5d})
\begin{eqnarray}
S=\int d^5x\,\sqrt{-g} \,\left[\frac{R}{2}-
\frac{1}{2}\partial_{\mu}\phi\partial^{\mu}\phi-U(\phi)
-\sum_{i=1}^{N}\frac{1}{4}\,e^{\lambda_{i}\phi}F_{(i) \mu
\nu}F_{(i)}^{\mu \nu}\right],
\end{eqnarray}
where the potential is
\begin{eqnarray}
 U(\phi)=\Lambda e^{\gamma
\phi}. \label{pot}
\end{eqnarray}
The equations of motions read
\begin{subequations}
\label{eqmotionghvm}
    \begin{eqnarray}
&&G_{\mu \nu}=T_{\mu\nu}-g_{\mu\nu}U(\phi),\\
\nonumber\\ &&\nabla_{\mu} \left(e^{\lambda_{i} \phi} F_{(i)}^{\mu
\nu}\right)=0,\label{eqmotionAmuhvm}\\
\nonumber\\ &&\Box \phi= \sum_{i=1}^{N}\left(\frac{\lambda_{i}}{4}
e^{\lambda_{i} \phi} F_{(i) \sigma \rho} F_{(i)}^{\sigma
\rho}\right)+\frac{d U}{d \phi},\label{eqmotionphihvm}
\end{eqnarray}
\end{subequations}
where the energy-momentum tensor $T_{\mu\nu}$ is given by
(\ref{tmn}). In what follows, we will present two dyonic solutions
of the field equations (\ref{eqmotionghvm}) with Solv and Nil's
horizon topologies.

%%%%%%%%%%%%%%%%%%%%%%%%%%%%%%%%%%%%%%%%%%%%%%%%%%%%%%%%%%%%%%%%%%%%%%%%%%%%
\subsection{Hyperscaling violation dyonic Solv's solution.}\label{Solvgemdyonic}
%%%%%%%%%%%%%%%%%%%%%%%%%%%%%%%%%%%%%%%%%%%%%%%%%%%%%%%%%%%%%%%%%%%%%%%%%%%%
We first report a solution of the field equations
(\ref{eqmotionghvm}) where the event horizon is given by the Solv's
geometry (\ref{Solvg}). The metric element and fields are given by
\begin{eqnarray}
\label{solv-solution-dyonic-hvm} ds^2&=& \frac{1}{r^{\frac{2
\theta}{3}}}\,\left[-{r}^{2z}\,f(r)dt^2+ \frac{1}{r^{2} f(r)} {dr^2}
+r^{2} x_3^{2}\,dx_1^2+r^{2}\,\frac{dx_2^2}{x_3^{2}}
+\left(\frac{2}{z+2-\theta}\right)\,\frac{dx_3^2}{x_3^{2}}\right],\nonumber\\
 \nonumber\\
f(r)&=&1-m\left(\frac{r_h}{r}\right)^{z+2-\theta}
+(m-1)\left(\frac{r_h}{r}\right)^{2z+2-2\theta},
\label{sol1bsolvdynhvm}\\ \nonumber\\
F_{(1)rt}&=&\sqrt {(z+2-\theta)(z-1)}\,{r}^{z+1-\theta},\quad
F_{(2)rt}=\sqrt {(z-\theta) (1-\theta) (m-1)} \left(\frac{r_h}{r}\right)^{z+1-\theta},\nonumber\\
F_{(3)x_1 x_2}&=&\sqrt {(z-\theta)(m-1)}\, {r_h}^{z+1-\theta},\quad
e^{\phi}= r^{\sqrt{{2 (z-1)-\frac{\theta}{3}\, \left( 3\,z-\theta
\right)}}},\nonumber
 \label{sol1csolvedyonhvm}
\end{eqnarray}
provided that the coupling are tied as follows
\begin{eqnarray*}
\label{lambdasolvdyn2} \Lambda&=&-\frac{(z+2-\theta)^{2}}{2},\qquad
\gamma=\frac{2 \theta}{ \sqrt{18 (z-1)-3\theta (3\,z-\theta)}},\quad
\lambda_{1}=-\frac{4 (3- \theta)}{\sqrt{18 (z-1)-3\theta
(3\,z-\theta)}},\nonumber\\
\lambda_2&=&\frac{2(3\,z-3-\theta)}{\sqrt{18 (z-1)-3\theta
(3\,z-\theta)}},\quad \lambda_3=-\frac{2(3\,z-3-2\,\theta)}{\sqrt{18
(z-1)-3\theta (3\,z-\theta)}}.
\end{eqnarray*}
It is interesting to note that in this limiting case $\theta\to 0$,
the constant $\gamma$ goes to zero and hence the potential term
(\ref{pot}) becomes a cosmological constant term. Consequently, in
the absence of the hyperscaling violation factor $\theta=0$, the
solution reduces to the dyonic solution found previously, see Sec.
$3$.

As before, in order to provide a complete thermodynamics analysis of
the Solv's solution with hyperscaling violation, we opt for the
Hamiltonian formalism where the reduced action (\ref{redSolvaction})
becomes
\begin{eqnarray}
\label{redSolvactionhvm} && I_E=\beta \vert
\Omega_{\tiny{\mbox{Solv}}}\vert \int _{r_h}^{\infty} \left(N{\cal
H}+\sum_{i=1}^2A_{(i)\tau}{\cal
\bar{P}}_{(i)}^{\prime} \right)dr+B_E,\\
&& {\cal
H}=\sqrt{\frac{2}{z+2-\theta}}\Bigg[r^{2-\theta}\left(1+\frac{z-\theta}{2}\right)+
\frac{1}{r^{\theta}}F\left(1+\frac{\theta^2-3\theta}{3}\right)+r^{1-\theta}F^{\prime}\left(1-\frac{\theta}{2}\right)+
\frac{r^{2-\theta}}{2} F (\phi^{\prime})^2\nonumber\\
&&\qquad\qquad\qquad\qquad +\sum_{i=1}^2
\frac{(z+2-\theta)}{4r^{2-\frac{\theta}{3}}}e^{-\lambda_i\phi}{\cal
\bar{P}}_{(i)}^2+
\frac{e^{\lambda_3\phi}}{2\,r^{2+\frac{\theta}{3}}}(F_{(3)x_1x_2})^2\Bigg]\nonumber.
\end{eqnarray}
and where the conjugate momenta are given by
$$
{\cal
\bar{P}}_{(i)}=\sqrt{\frac{2}{z+2-\theta}}\,\frac{r^{2-\frac{\theta}{3}}}{N}\,e^{\lambda_i\phi}\,\partial_r
A_{(i)\tau},\qquad\qquad i=1, 2.
$$
The variation of the boundary term yields
\begin{eqnarray*}
\delta B_E=\beta \vert \Omega_{\tiny{\mbox{Solv}}}\vert\Bigg[&&
\sqrt{\frac{2}{z+2-\theta}}\,\left(\delta\left(\Big(1-\frac{\theta}{2}\Big)mr_h^{z+2-\theta}\right)-
\frac{2\pi}{\beta}\delta
r_h^{2-\theta}\right)\\
&&-\Phi_e\,\delta\left(\sqrt{\frac{2(z-\theta)(1-\theta)(m-1)}{z+2-\theta}}\,r_h^{z+1-\theta}\right)-
\sqrt{\frac{2(m-1)}{z+2-\theta}}\,r_h\,\delta\left(\sqrt{m-1}\,r_h^{z+1-\theta}\right)\Bigg],
\end{eqnarray*}
with
$$
\Phi_e=\sqrt{\frac{(1-\theta)(m-1)}{(z-\theta)}}\,r_h.
$$
After some computations, we finally conclude that the
thermodynamical quantities are
\begin{eqnarray}
&&{\cal M}=\vert \Omega_{\tiny{\mbox{Solv}}}\vert
\sqrt{\frac{2}{z+2-\theta}}\left(1-\frac{\theta}{2}\right)\,m\,
r_h^{z+2-\theta},\qquad T=\frac{\left[\left( m-2 \right) \theta-z
\left( m-2 \right) +2\right] r_h^{z}}{4 \pi},\nonumber\\
&&{\cal S}=2\pi \vert \Omega_{\tiny{\mbox{Solv}}}\vert
 \sqrt{\frac{2}{z+2-\theta}}\,r_h^{2-\theta},\qquad \Phi_e=\sqrt{\frac{(1-\theta)(m-1)}{(z-\theta)}}\,r_h,\nonumber\\
\nonumber\\
&&{\cal Q}_e=\vert \Omega_{\tiny{\mbox{Solv}}}
\vert\sqrt{\frac{2(z-\theta)(1-\theta)(m-1)}{z+2-\theta}}\,r_h^{z+1-\theta},\qquad
\Phi_m=A\sqrt{m-1}\,r_h,\\
\nonumber\\
&&{\cal Q}_m=\vert \Omega_{\tiny{\mbox{Solv}}}\vert
B\,\sqrt{m-1}\,r_h^{z+1-\theta},\qquad
AB=\sqrt{\frac{2}{z+2-\theta}},\nonumber \label{thermoqtesSolvhyp}
\end{eqnarray}
and we check again the validity of the first law (\ref{1stlawSolv}).

To end this section, we would like to point out an interesting
observation. For the Solv's solution without hyperscaling violation
parameter, we have shown that the charged solution must necessarily
be electric and magnetic. Here, the presence of the hyperscaling
factor $\theta$ allows to switch off the electric contribution
putting $\theta=1$, and the resulting configuration turns to be
purely magnetic.

%%%%%%%%%%%%%%%%%%%%%%%%%%%%%%%%%%%%%%%%%%%%%%%%%%%%%%%%%%%%%%%%%%%%%%%%%%%%
\subsection{Hyperscaling violation black hole with a Nil's Geometry}\label{Nilgemdyonic}
%%%%%%%%%%%%%%%%%%%%%%%%%%%%%%%%%%%%%%%%%%%%%%%%%%%%%%%%%%%%%%%%%%%%%%%%%%%%
We now turn to the construction of the dyonic extension of the
vacuum Nil's solution \cite{Hassaine:2015ifa} which is given by
\begin{eqnarray}
ds^2&=& \frac{1}{r^{\frac{2 \theta}{3}}}\,\left[-{r}^{2z}\,f(r)dt^2+
\frac{1}{r^{2} f(r)} dr^{2}+r^{2} dx_1^2+r^{2} dx_2^2
+\left(z+4-\theta\right) r^{4} (dx_3-x_1dx_2)^2\right], \nonumber\\
f(r)&=&1-m\left(\frac{r_h}{r}\right)^{z+4-\theta}
+(m-1)\left(\frac{r_h}{r}\right)^{2z+4-2 \theta},\quad e^{\phi}=
r^{\sqrt{2(2z-3)-\frac{\theta}{3}\left( 3\,z-\theta \right) }},
\label{sol1bnildyon}\nonumber\\ \\
 F_{(1) rt}&=&\frac{\sqrt {2
\,(2\,z-3)(z+4-\theta)}}{2}\,{{r}^{z+3-\theta}}, \qquad
F_{(2)rt}=\frac{\sqrt {(2-\theta)\, (z-\theta) (m-1)
}r_h^{z+2-\theta}}
 {r^{z+1-\theta}},\nonumber\\
F_{(3)x_1 x_2}&=&-x_1\sqrt{ (z-\theta) \left( m-1 \right) \left(
z+4-\theta \right)}\,r_h^{z+2-\theta}, \nonumber\\
F_{(3)x_1
x_3}&=&\sqrt{ (z-\theta) \left( m-1 \right) \left(
z+4-\theta \right)}\,r_h^{z+2-\theta},\nonumber\\
F_{(4)x_2 x_3}&=&\sqrt{ (z-\theta) \left( m-1 \right) \left(
z+4-\theta \right)}\,r_h^{z+2-\theta},\nonumber
\end{eqnarray}
provided that
\begin{eqnarray*}
\Lambda&=&-\frac { \left( z+4 -\theta \right)  \left( z+3-\theta
\right) }{2},\quad \lambda_1=-\frac{4(6-\theta)}{\sqrt{18 (2z-3)
-3\,\theta\, \left(
3\,z-\theta \right) }},\nonumber\\
\lambda_2&=&\frac{2\,(3\,z-6-\theta)}{\sqrt{18 (2z-3) -3\,\theta\,
\left( 3\,z-\theta \right) }},\quad \lambda_3=\lambda_4=-\frac{2
(3\,z-2\,\theta-3)}{\sqrt{18 (2z-3) -3\,\theta\, \left( 3\,z-\theta
\right) }},
\end{eqnarray*}
and the Liouville coupling potential (\ref{pot}) is given by
$$
\gamma=\frac{2 \theta}{ \sqrt{18 (2z-3) -3\,\theta\, \left(
3\,z-\theta \right) }}.
$$

In the limiting case, $z=3/2$, $\theta=9/2$ and $m=1$, one
effectively recovers the vacuum solution found in
\cite{Hassaine:2015ifa}. Also, note that for a hyperscaling
violation factor $\theta=2$, the solution can be rendered purely
magnetic. Now, proceeding as before, one obtains the following
thermodynamical quantities
\begin{eqnarray}
&&{\cal M}=\vert \Omega_{\tiny{\mbox{Nil}}}\vert
\sqrt{z+4-\theta}\,\left(2-\frac{\theta}{2}\right)\,m\,
r_h^{z+4-\theta},\qquad
T=\frac{\left[\left( m-2 \right) \theta+z(2-m)+4\right] r_h^{z}}{4 \pi}, \nonumber\\
&&{\cal S}=2\pi \vert \Omega_{\tiny{\mbox{Nil}}}\vert
\sqrt{z+4-\theta}\,r_h^{4-\theta},
\quad \Phi_e=\sqrt{\frac{(2-\theta)(m-1)}{z-\theta}}\,r_h^2,\nonumber\\
 &&{\cal
Q}_e=\vert \Omega_{\tiny{\mbox{Nil}}}\vert
\sqrt{(z+4-\theta)(2-\theta)(m-1)(z-\theta)}\,r_h^{z+2-\theta},\qquad \Phi_m=A\sqrt{m-1}\,r_h^2,\nonumber\\
\\
 &&{\cal Q}_m=\vert
\Omega_{\tiny{\mbox{Nil}}}\vert\,B\sqrt{m-1}\,r_h^{z+2-\theta},\qquad
AB=2\sqrt{z+4-\theta}, \nonumber \label{thermoqtesNilhyp}
\end{eqnarray}
which satisfy the electromagnetic version of the first law of
thermodynamics (\ref{1stlawSolv}).

%%%%%%%%%%%%%%%%%%%%%%%%%%
\section{Conclusions}
%%%%%%%%%%%%%%%%%%%%%%%%%%
Here, we have shown that the vacuum solutions with Solv and Nil's
horizon topologies of the five-dimensional Einstein equations can be
electromagnetically charged through a dilatonic source with at least
three Abelian gauge fields. The resulting dyonic solutions are
asymptotically anisotropic and can be considered as Lifshitz black
holes in the sense as defined by Eqs.
(\ref{Lifshitznonst}-\ref{aniTransgener}). The presence of various
Abelian fields is mandatory in order to ensure the Lifshitz
asymptotic and the emergence of the electric and magnetic charges.
Through an Hamiltonian approach, we have realized a complete
analysis of the thermodynamics features of the dyonic solutions, and
we have checked that for each solution, the electromagnetic version
of the first law of thermodynamics is satisfied. We have noticed
that the dyonic solutions, in spite of having a
Reissner-Nordstrom-like metric, are quite different from the
magnetically charged Reissner-Nordstrom solution. Indeed, for the
dyonic Lifshitz solutions with Thurston's horizon topologies, the
electric and magnetic charges turn to be proportional. In other
words, there does not exist a purely electric or purely magnetic
solution. This characteristic is similar to what occur for the
odd-dimensional Chern-Simons vortices (see \cite{Dunne:1998qy} for a
good review). Indeed, because of the presence of the
three-dimensional Chern-Simons term $\kappa
\epsilon^{\mu\nu\rho}A_{\mu}F_{\nu\rho}$ in the action, the
magnetically charged vortices also carry an electric charge
proportional to the magnetic charge. It is important to stress again
that the presence of the dyonic charges allows the Lifshitz
dynamical exponent to be free and not restricted as in the vacuum
case. Such feature was already observed in \cite{Alvarez:2014pra}
where the presence of a nonlinear electrodynamics source was
responsible of the freedom of the dynamical exponent.

The hyperscaling violation extensions of these dyonic solutions were
also considered. In this case,  the dilatonic source is augmented by
a Liouville potential term and the cosmological constant is turned
off. We have noticed that for some specific values of the
hyperscaling violation factor, the dyonic solutions can be rendered
purely magnetic.

An interesting work to be done will consist in computing for the
dyonic solutions reported here  the DC conductivities of the
corresponding field theory in order to gain some precision about
this latter, see e. g. \cite{Arias:2017yqj}. Also, very recently, a
new dyonic solution of the Einstein-Maxwell-dilaton theory was
constructed in \cite{Stelea:2017pdk} using sone solution-generating
technique. It will be interesting to see wether these techniques can
be exported in our problem to generate news dyonic solutions.

%%%%%%%%%%%%%%%%%%%%%%%%%%%


\begin{thebibliography}{99}
%%%%%%%%%%%%%%%%%%%%%%%%%%%
\bibitem{Son:2008ye}
 D.~T.~Son,
  %``Toward an AdS/cold atoms correspondence: A Geometric realization of the Schrodinger symmetry,''
  Phys.\ Rev.\ D {\bf 78}, 046003 (2008).
 %%CITATION = doi:10.1103/PhysRevD.78.046003;%%


\bibitem{Balasubramanian:2008dm}
  K.~Balasubramanian and J.~McGreevy,
  %``Gravity duals for non-relativistic CFTs,''
  Phys.\ Rev.\ Lett.\  {\bf 101}, 061601 (2008)
%%CITATION = doi:10.1103/PhysRevLett.101.061601;%%

\bibitem{Duval:2008jg}
  C.~Duval, M.~Hassaine and P.~A.~Horvathy,
  %``The Geometry of Schrodinger symmetry in gravity background/non-relativistic CFT,''
  Annals Phys.\  {\bf 324}, 1158 (2009)
%%CITATION = doi:10.1016/j.aop.2009.01.006;%%


\bibitem{Kachru:2008yh}
  S.~Kachru, X.~Liu and M.~Mulligan,
  %``Gravity duals of Lifshitz-like fixed points,''
  Phys.\ Rev.\ D {\bf 78}, 106005 (2008).
%%CITATION = doi:10.1103/PhysRevD.78.106005;%%


\bibitem{Taylor:2008tg}
  M.~Taylor,``Non-relativistic holography,'' arXiv:0812.0530 [hep-th].
%%CITATION = ARXIV:0812.0530;%%

\bibitem{AyonBeato:2009nh}
  E.~Ayon-Beato, A.~Garbarz, G.~Giribet and M.~Hassaine,
  %``Lifshitz Black Hole in Three Dimensions,''
  Phys.\ Rev.\ D {\bf 80}, 104029 (2009); JHEP {\bf 1004}, 030
  (2010).
%%CITATION = ARXIV:0909.1347
%%CITATION = doi:10.1007/JHEP04(2010)030;%%


\bibitem{Pang:2009pd}
  D.~-W.~Pang,
  %``On Charged Lifshitz Black Holes,''
  JHEP {\bf 1001}, 116 (2010).
%%CITATION = ARXIV:0911.2777;%%

\bibitem{Maeda:2011jj}
  H.~Maeda and G.~Giribet,
  %``Lifshitz black holes in Brans-Dicke theory,''
  JHEP {\bf 1111}, 015 (2011).
%%CITATION = ARXIV:1105.1331;%%



\bibitem{Zangeneh:2016cbv}
  M.~K.~Zangeneh, A.~Dehyadegari, A.~Sheykhi and M.~H.~Dehghani,
  %``Thermodynamics and gauge/gravity duality for Lifshitz black holes in the presence of exponential electrodynamics,''
  JHEP {\bf 1603}, 037 (2016).
  %%CITATION = doi:10.1007/JHEP03(2016)037;%%

\bibitem{Quinta:2016eql}
  G.~M.~Quinta, A.~Flachi and J.~P.~S.~Lemos,
  %``Vacuum polarization in asymptotically Lifshitz black holes,''
  Phys.\ Rev.\ D {\bf 93}, no. 12, 124073 (2016).
  %%CITATION = doi:10.1103/PhysRevD.93.124073;%%



\bibitem{Ayon-Beato:2015jga}
  E.~Ayon-Beato, M.~Bravo-Gaete, F.~Correa, M.~Hassaine, M.~M.~Juarez-Aubry and J.~Oliva,
  %``First law and anisotropic Cardy formula for three-dimensional Lifshitz black holes,''
  Phys.\ Rev.\ D {\bf 91}, no. 6, 064006 (2015).
%%CITATION = ARXIV:1501.01244;%%

\bibitem{Bravo-Gaete:2015xea}
  M.~Bravo-Gaete and M.~Hassaine,
  %``Thermodynamics of charged Lifshitz black holes with quadratic corrections,''
  Phys.\ Rev.\ D {\bf 91}, no. 6, 064038 (2015).
%%CITATION = ARXIV:1501.03348;%%


\bibitem{Zangeneh:2015uwa}
  M.~Kord Zangeneh, A.~Sheykhi and M.~H.~Dehghani,
  %``Thermodynamics of topological nonlinear charged Lifshitz black holes,''
  Phys.\ Rev.\ D {\bf 92}, no. 2, 024050 (2015).
%%CITATION = doi:10.1103/PhysRevD.92.024050;%%


\bibitem{Correa:2014ika}
  F.~Correa, M.~Hassaine and J.~Oliva,
  %``Black holes in New Massive Gravity dressed by a (non)minimally coupled scalar field,''
  Phys.\ Rev.\ D {\bf 89}, no. 12, 124005 (2014).
%%CITATION = ARXIV:1403.6479;%%


\bibitem{Fan:2015yza}
  Z.~Y.~Fan and H.~Lu,
  %``Charged Black Holes in Colored Lifshitz Spacetimes,''
  Phys.\ Lett.\ B {\bf 743}, 290 (2015).
%%CITATION = doi:10.1016/j.physletb.2015.02.052;%%


\bibitem{Hendi:2013zba}
  S.~H.~Hendi, B.~Eslam Panah and C.~Corda,
  %``Asymptotically Lifshitz black hole solutions in F(R) gravity,''
  Can.\ J.\ Phys.\  {\bf 92}, no. 1, 76 (2014).
  %%CITATION = doi:10.1139/cjp-2013-0357;%%


\bibitem{Alvarez:2014pra}
  A.~Alvarez, E.~Ayon-Beato, H.~A.~Gonzalez and M.~Hassaine,
  %``Nonlinearly charged Lifshitz black holes for any exponent $z>1$,''
  JHEP {\bf 1406}, 041 (2014).
%%CITATION = ARXIV:1403.5985;%%


\bibitem{Bravo-Gaete:2013dca}
  M.~Bravo-Gaete and M.~Hassaine,
  %``Lifshitz black holes with a time-dependent scalar field in a Horndeski theory,''
  Phys.\ Rev.\ D {\bf 89}, 104028 (2014).
%%CITATION = doi:10.1103/PhysRevD.89.104028;%%




\bibitem{Cadeau:2000tj}
  C.~Cadeau and E.~Woolgar, Class.\ Quant.\ Grav.\  {\bf 18}, 527 (2001).
  %%CITATION = GR-QC/0011029;%

\bibitem{Mann:2009yx}
  R.~B.~Mann,
  %``Lifshitz Topological Black Holes,''
  JHEP {\bf 0906}, 075 (2009).
%%CITATION = doi:10.1088/1126-6708/2009/06/075;%%


\bibitem{Matulich:2011ct}
  J.~Matulich and R.~Troncoso,
  %``Asymptotically Lifshitz wormholes and black holes for Lovelock gravity in vacuum,''
  JHEP {\bf 1110}, 118 (2011).
%%CITATION = doi:10.1007/JHEP10(2011)118;%%



\bibitem{Olivares:2013uha}
  M.~Olivares, G.~Rojas, Y.~Vasquez and J.~R.~Villanueva,
  %``Particles motion on topological Lifshitz black holes in 3+1 dimensions,''
  Astrophys.\ Space Sci.\  {\bf 347}, 83 (2013).
%%CITATION = doi:10.1007/s10509-013-1496-0;%%



\bibitem{Hassaine:2015ifa}
  M.~Hassaine,
  %``New black holes of vacuum Einstein equations with hyperscaling violation and Nil geometry horizons,''
  Phys.\ Rev.\ D {\bf 91}, no. 8, 084054 (2015).
%%CITATION = doi:10.1103/PhysRevD.91.084054;%%


%\cite{Tarrio:2011de}
\bibitem{Tarrio:2011de}
  J.~Tarrio and S.~Vandoren,
  %``Black holes and black branes in Lifshitz spacetimes,''
  JHEP {\bf 1109}, 017 (2011).
%%CITATION = doi:10.1007/JHEP09(2011)017;%%



\bibitem{Hervik:2003vx}
  S.~Hervik,
  %``Einstein metrics: Homogeneous solvmanifolds, generalized Heisenberg groups and black holes,''
  J.\ Geom.\ Phys.\  {\bf 52}, 298 (2004).
%%CITATION = doi:10.1016/j.geomphys.2004.03.005;%%

\bibitem{Thurston}
 W. P. Thurston {\it Three-Dimensional Geometry and Topology},
ed. S. Levy (Princeton University Press, Princeton 1997).


\bibitem{Liu:2014dva}
  H.~S.~Liu and H.~Lü,
  %``Thermodynamics of Lifshitz Black Holes,''
  JHEP {\bf 1412}, 071 (2014).
%%CITATION = doi:10.1007/JHEP12(2014)071;%%



\bibitem{Babichev:2017rti}
  E.~Babichev, C.~Charmousis and M.~Hassaine,
  %``Black holes and solitons in an extended Proca theory,''
  JHEP {\bf 1705}, 114 (2017).
%%CITATION = doi:10.1007/JHEP05(2017)114;%%


\bibitem{Dehghani}
  M.~H.~Dehghani, A.~Sheykhi and S.~H.~Hendi,
  %``Magnetic Strings in Einstein-Born-Infeld-Dilaton Gravity,''
  Phys.\ Lett.\ B {\bf 659}, 476 (2008).
%%CITATION = doi:10.1016/j.physletb.2007.11.015;%%


\bibitem{Hendi:2015cra}
  S.~H.~Hendi, M.~Faizal, B.~E.~Panah and S.~Panahiyan,
  %``Charged dilatonic black holes in gravity's rainbow,''
  Eur.\ Phys.\ J.\ C {\bf 76}, no. 5, 296 (2016).
%%CITATION = doi:10.1140/epjc/s10052-016-4119-4;%%


\bibitem{Hendi:2017phi}
  S.~H.~Hendi, N.~Riazi, S.~Panahiyan and B.~Eslam Panah,
  ``Higher dimensional dyonic black holes,'' arXiv:1710.01818 [gr-qc].
%%CITATION = ARXIV:1710.01818;%%


\bibitem{Arias:2017yqj}
R.~E.~Arias and I.~S.~Landea,``Thermoelectric Transport Coefficients
from Charged Solv and Nil Black Holes,'' arXiv:1708.04335 [hep-th].
%%CITATION = ARXIV:1708.04335;%%



\bibitem{Gibbons:1976ue}
  G.~W.~Gibbons and S.~W.~Hawking,
  %``Action Integrals and Partition Functions in Quantum Gravity,''
  Phys.\ Rev.\ D {\bf 15}, 2752 (1977).
%%CITATION = doi:10.1103/PhysRevD.15.2752;%%





\bibitem{Cardenas:2016uzx}
  M.~Cardenas, O.~Fuentealba and J.~Matulich,
  %``On conserved charges and thermodynamics of the AdS$_{4}$ dyonic black hole,''
  JHEP {\bf 1605}, 001 (2016).
%%CITATION = doi:10.1007/JHEP05(2016)001;%%




\bibitem{Charmousis:2012dw}
  C.~Charmousis, B.~Gouteraux and E.~Kiritsis, JHEP {\bf 1209}, 011 (2012).
%%CITATION = ARXIV:1206.1499;%%


\bibitem{Dong:2012se}
  X.~Dong, S.~Harrison, S.~Kachru, G.~Torroba and H.~Wang,
  %``Aspects of holography for theories with hyperscaling violation,''
  JHEP {\bf 1206}, 041 (2012).
%%CITATION = ARXIV:1201.1905;%%

\bibitem{Alishahiha:2012qu}
  M.~Alishahiha, E.~O Colgain and H.~Yavartanoo,
  %``Charged Black Branes with Hyperscaling Violating Factor,''
  JHEP {\bf 1211}, 137 (2012).
%%CITATION = ARXIV:1209.3946;%%


\bibitem{HVMBH1} M.~Cadoni and M.~Serra, JHEP {\bf 1211}, 136 (2012).
%%CITATION = ARXIV:1209.4484;%%

\bibitem{HVMBH2} P.~Bueno, W.~Chemissany, P.~Meessen, T.~Ortin and C.~S.~Shahbazi,
  %``Lifshitz-like Solutions with Hyperscaling Violation in Ungauged Supergravity,''
  JHEP {\bf 1301}, 189 (2013).
%%CITATION = ARXIV:1209.4047;%%



\bibitem{Dunne:1998qy}
  G.~V.~Dunne, ``Aspects of Chern-Simons theory,'' hep-th/9902115.
%%CITATION = HEP-TH/9902115;%%

\bibitem{Stelea:2017pdk}
  C.~Stelea, ``New solutions of the Einstein-Maxwell-Dilaton theory in five
  dimensions'',
  arXiv:1709.10309 [gr-qc].
  %%CITATION = ARXIV:1709.10309;%%





\end{thebibliography}
\end{document}